\documentclass[twocolumn,aps,prl,superscriptaddress]{revtex4}
\usepackage{amsfonts}
\usepackage{amsmath}
\usepackage{amssymb}
\usepackage{graphicx}
\usepackage{color}
\usepackage{ulem}
\usepackage[colorlinks, urlcolor=cyan, citecolor=blue, linkcolor=magenta]{hyperref}

\begin{document}

\title{Topological Origin of Floquet Thermalization \\
in Periodically Driven Many-body Systems}
\author{Hao-Yue Qi}
\thanks{These authors contribute equally to this work.}
\affiliation{Hefei National Research Center for Physical Sciences at the Microscale and
School of Physical Sciences, University of Science and Technology of China,
Hefei 230026, China}
\affiliation{CAS Center for Excellence in Quantum Information and Quantum Physics,
University of Science and Technology of China, Hefei 230026, China}
\author{Yue Wu}
\thanks{These authors contribute equally to this work.}
\affiliation{Institute for Advanced Study, Tsinghua University, Beijing 100084, China}
\author{Wei Zheng}
\email{zw8796@ustc.edu.cn}
\affiliation{Hefei National Research Center for Physical Sciences at the Microscale and
School of Physical Sciences, University of Science and Technology of China,
Hefei 230026, China}
\affiliation{CAS Center for Excellence in Quantum Information and Quantum Physics,
University of Science and Technology of China, Hefei 230026, China}
\affiliation{Hefei National Laboratory, University of Science and Technology of China,
Hefei 230088, China}
\date{\today }

\begin{abstract}
Floquet engineering is a powerful manipulation method in modern quantum technology. However, unwanted heating is the main challenge of Floquet engineering, therefore the Floquet thermalization has attracting considerable attentions recently.
In this work, we investigate thermalization of periodically driven many-body
systems through the lens of Krylov complexity, and find a topological origin of different thermalization behaviors.
We demonstrate that If the
topology of the Krylov chain is nontrivial, a periodically driven system will reach a state
with finite temperature. When the Krylov chain is topologically trivial, the
system will be heated to infinite temperature. We further show that
the prethermalization can be understood as the tunnelling process
of a quasi-edge mode through the local gap on Krylov chain. This picture
provides a systematically method to obtain the effective prethermal Hamiltonian.
\end{abstract}

\date{\today}
\maketitle

Quantum systems with periodic driving have attracted much attention in the
decades. First, periodic driving can be used to engineer nontrivial
Hamiltonians. For example, periodic driving has been used to generate
artificial magnetic fields in optical lattice, simulate synthetic lattice
gauge theories with ultracold atomic gases~\cite{Bloch@2019.Floquet-LGT},
and engineer topological bands~\cite%
{Zhai@2014.Floquet-TP,Esslinger@2014.Floquet-Haldane}. Second, periodic
driving can induce intrinsically out-of-equilibrium phenomena without
counterpart in equilibrium, such as discrete time-crystals~\cite%
{Yao@2023.DTC-REV} and anomalous Floquet topological states~\cite%
{Demler@2010.Anomalous-Floquet,Levin@2013.Floquet-TP,Bloch@2020.Anomalous-Floquet,Zhai@2022.TP-MicroMotion}%
.However, since periodic driving breaks energy conservation, it was expected
that under driving, generic many-body systems would be heated up to featureless
infinite temperature ensembles eventually~\cite%
{Rigol@2014.Floquet-ETH,Moessner@2014.Floquet-ETH,Huse@2014.F-Heating,Papic@2015.F-Heating}%
.
Fortunately, under high-frequency driving, there is a class of systems that
exhibit the prethermalization phenomena~\cite%
{Huveneers@2015.PreThermalization,Saito@2016.PreThermalization,Kuwahara@2016.PreTh,Abanin@2017.PreTH, Bukov@2016.PreThermalization,Weidinger@2016.PreThermalization,Canovi@2016.PreThermalization, Abanin@2017.PreThermalization,Bukov@2021.PreTH,Ho@2023.PreThermalization-REV}%
. Such systems have several approximate conservation quantities. Thus, at
intermediate time scales, systems will first equilibrate to a finite
temperature state governed by a so-called prethermal Hamiltonian. Later,
systems will thermalize to infinite temperature ensembles at much longer
time scales. Prethermalization has been experimentally observed in one-dimensional Bose gas and spins with dipole interaction ~\cite%
{Bloch@2020.PreTH-EXP,Peng@2021.PreThermalization,Beatrez@2021.PreTH-EXP}. However, in what circumstances, a quantum system will be directly heated to infinite temperature or exhibit prethermalization behavior is not clear. There is no simple and clear physical picture of the Floquet thermalization.

Recently, the Krylov complexity of operators has been developed to study the
chaos and thermalization in many-body systems~\cite{Altman@2019.Krylov}.
With the help of Krylov basis, the evolution of an operator under
Heisenberg equation can be mapped to the spreading of a wavepacket in a one-dimensional tight-binding chain. 
Krylov complexity has motivated a lot of interesting studies in closed~\cite{Zhou@2023.Krylov,Zhai@2023.Krylov,Dymarsky@2021.Krylov,Rabinovici@2022.Krylov,Caputa@2022.Krylov,
Jian@2023.Krylov,Dymarsky@2020.Krylov,Mitra@2020.Krylov,Mitra@2023.Krylov,ZhangPF@2023.Krylov,ZhangPF@2024.Krylov}
\cite{Yates@2021.Krylov-edge,Mitra@2022.F-Krylov,Shrestha@2023.F-Krylov,Suchsland@2023.Krylov, Mitra@2023.F-Krylov}
and open~\cite%
{Zhai@2023.Open-Krylov,Bhattacharya@2023.Krylov-Open,Srivatsa@2023.Krylov-Open}
quantum many-body systems.

In this work, we found topological origin of different thermalization behavior of periodically driven many-body systems based on Krylov complexity.
We first build time-independent Krylov basis for periodically driven systems, and map the Floquet dynamics of an operator to the wavepacket evolution on the Krylov chain. The main results of this work are summarized here:

(1) We construct the unfold Floquet effective Hamiltonian by seeking the zero mode on Krylov chain.
Unlike the effective Hamiltonian obtained by taking the logarithm of evolution operator in previous works, which has folded spectrum and does not satisfy the assumption of eigenstate thermalization hypothesis (ETH), our unfold effective Hamiltonian satisfies ETH and can be used
to estimate local observables at long time scale.

(2) We further find that it is the topology of Krylov chain that determined the Floquet thermalization behavior.
In a finite system, when the Krylov chain is topologically nontrivial,
it sustains a localized zero edge mode. Then the system will heated to a state with finite
temperature. If topology of the Krylov chain is trivial, the zero mode is
extended to the bulk. As a result, the system will be heated to infinite
temperature.

(3) In the thermodynamic limit, the Krylov chain is
gapless. Therefore there is no eigen zero edge mode, and the system will
finally be heated to infinite temperature. However, we find that there could
exist a quasi edge mode in high driving frequency regime, which is protected by the local
 topology and gap of the Krylov chain. We further conjecture that this quasi edge mode represents the Floquet
prethermal Hamiltonian. Its life time determines the time scale of
intermediate prethermal state. With this picture in mind, one can systematically estimate the prethermal Hamiltonian via Krylov basis.

\begin{figure}[tbp]
\centering
\includegraphics[width=\columnwidth]{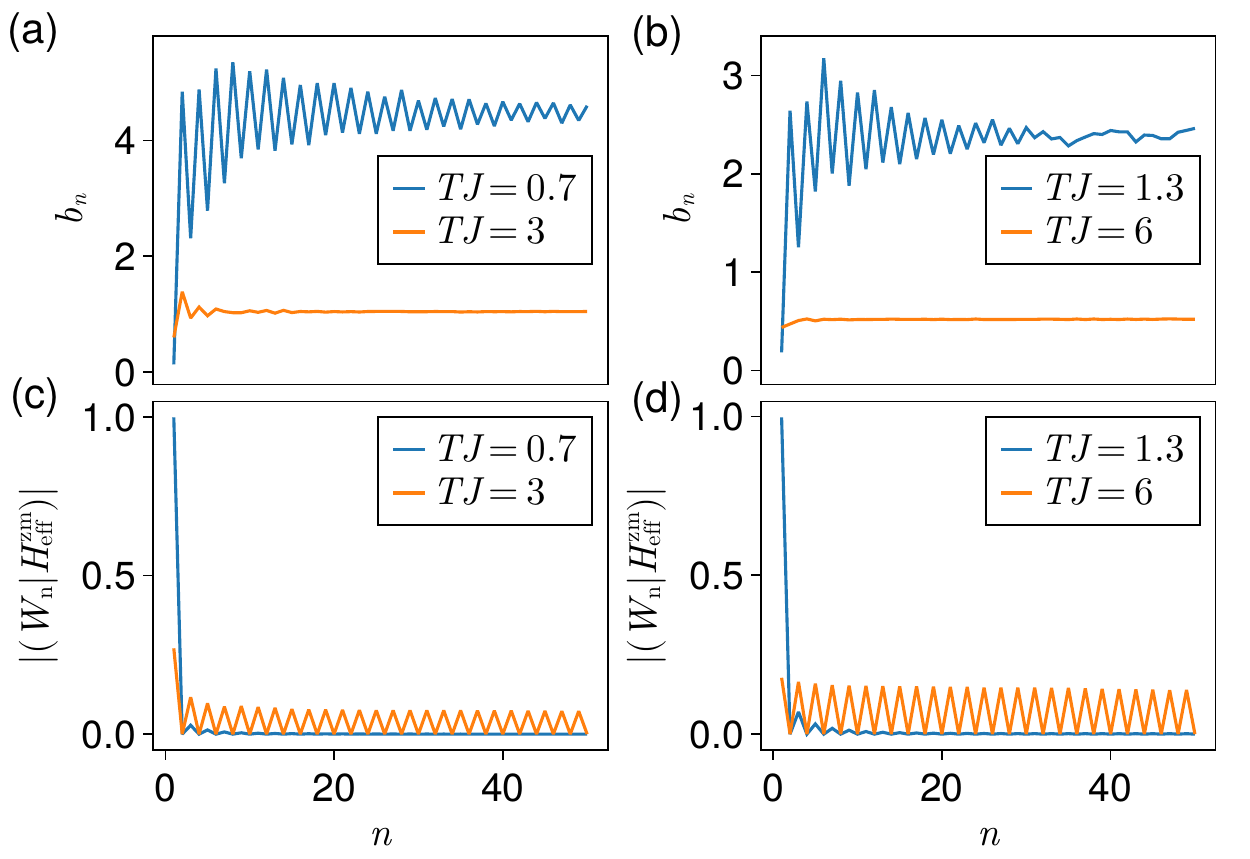}
\caption{Hopping amplitude $b_n$ and zero mode $|H_{\mathrm{eff}}^{\mathrm{zm%
}})$ on the Krylov chain. (a)(c) quantum spin chain with system size $L=14$
and $h_z/J = 0.809,h_x/J = 0.9045$. (b)(d) hard-core Bose-Hubbard model with
system size $L=16$ and $J^{\prime}/J = 0.2,U/J=1$. For both two models, the
hopping amplitude fluctuates around a non-zero value site-by-site in
high-frequency regime, and the fluctuation becomes smaller deep in the bulk.
The zero mode is an edge mode. In contrast, in the low-frequency regime,
the hopping amplitude is nearly constant. The zero mode extends over the
bulk.}
\label{fig1}
\end{figure}

\textit{Krylov Chain.} In a periodically driven system, $\hat{H}(t+T)=\hat{H}%
(t)$, the evolution operator during one driving period is given by $\hat{U}_{%
\mathrm{F}}=\mathcal{\hat{T}}\exp \left[ -i\int_{0}^{T}dt\hat{H}(t)\right] $%
, here we set $\hbar =1$. One can define a time-independent effective
Hamiltonian as%
\begin{equation}
\hat{U}_{\mathrm{F}}=e^{-i\hat{H}_{\mathrm{eff}}T}.  \label{Heff}
\end{equation}%
This effective Hamiltonian reproduces the evolution of the driven system in
a stroboscopic fashion. In other words, the state evolution under
time-independent $\hat{H}_{\mathrm{eff}}$ is identical to the one under
driven Hamiltonian $\hat{H}(t)$, only at steps of the driving period $t=nT$.
Thus the effective Hamiltonian well describes the dynamics over time scale
much longer than a single driving period. Numerically, we calculate the
effective Hamiltonian by taking the logarithm of $\hat{U}_{\mathrm{F}}$, as $%
\hat{H}_{\mathrm{eff}}^{\mathrm{fold}}=\frac{i}{T}\ln \hat{U}_{\mathrm{F}}$.
Therefore, the long-time evolution of operators can be captured by the
following Heisenberg equation in a stroboscopic way,
\begin{equation}
i\partial _{t}\hat{O}=\left[ \hat{O},\hat{H}_{\mathrm{eff}}^{\mathrm{fold}}%
\right] .  \label{HB-equ}
\end{equation}

However, the effective Hamiltonian obtained by this method is not unique. It
has a folded energy spectrum. Specifically speaking, $\hat{H}_{\mathrm{eff}%
}^{\mathrm{fold}}$ can be diagonalized into $\hat{H}_{\mathrm{eff}}^{\mathrm{%
fold}}=\sum\nolimits_{k}E_{k}\left\vert \psi _{k}\right\rangle \left\langle
\psi _{k}\right\vert $, and its eigen energies are folded in the first
Floquet-Brillouin zone, $-\pi /T<E_{k}<\pi /T$. As a result, it does not
satisfy the assumption of ETH, and can not predict the expectation values of
local observables after a long time evolution. Certainly, we can construct
other equivalent effective Hamiltonian by shifting the eigenenergies without
altering the eigenstates as $\hat{H}_{\mathrm{eff}}=\sum\nolimits_{k}\left(
E_{k}+m_{k}\Omega \right) \left\vert \psi _{k}\right\rangle \left\langle
\psi _{k}\right\vert $, where $m_{k}$ is an integer and $\Omega =2\pi /T$ is
the driving frequency. Such Hamiltonians satisfy the definition (\ref{Heff}%
), and thus commute with $\hat{H}_{\mathrm{eff}}^{\mathrm{fold}}$.

The Krylov basis of operators is based on the Lanczos algorithm. An operator
is denoted by $\hat{O}=\sum\nolimits_{ij}O_{ij}\left\vert i\right\rangle
\left\langle j\right\vert $, where $\{\left\vert i\right\rangle\}$ is a set of
orthogonal states and $O_{ij}$ are corresponding matrix elements. We can map
the operator $\hat{O}$ to a vector in the operator space as $\hat{O}%
\longrightarrow \left\vert O\right) =\sum\nolimits_{ij}O_{ij}\left\vert
i\right\rangle \otimes \left\vert j\right\rangle $. The corresponding inner
product of operators is defined as $\left( O_{1}|O_{2}\right) =\mathrm{Tr}(%
\hat{O}_{1}^{\dag }\hat{O}_{2})$. We further introduce a Liouville
super-operator $\mathcal{L}$, acting on an operator as $\mathcal{L}\hat{O}%
\equiv \lbrack \hat{H}_{\mathrm{eff}}^{\mathrm{fold}},\hat{O}]$. In the
operator space, it can be mapped into a matrix, $\mathcal{L}\hat{O}%
\rightarrow \mathcal{L}\left\vert O\right) $. Thus the Heisenberg equation (%
\ref{HB-equ}) can be expressed into $i\partial _{t}\left\vert O\right) =-%
\mathcal{L}\left\vert O\right) $. Applying the the Baker-Campbell-Hausdorff
formula, we expand the evolution of operator into $\left\vert O(t)\right)
=\sum\nolimits_{n}\frac{\left( -it\right) ^{n}}{n!}\mathcal{L}^{n}\left\vert
O\right) $. Here $\left\{ \mathcal{L}^{n}\left\vert O\right) \right\} $
forms a complete set of operator bases, but it is neither normalized nor
orthogonal. We then apply the Gram-Schmidt procedure to obtain a set of
orthogonalized and normalized basis, called Krylov basis as,
\begin{eqnarray}
\left\vert W_{0}\right) &=&\frac{1}{b_{0}}\left\vert O\right) , \\
\left\vert W_{1}\right) &=&\frac{1}{b_{1}}\mathcal{L}\left\vert W_{0}\right)
, \\
\left\vert W_{n}\right) &=&\frac{1}{b_{n}}\left[ \mathcal{L}\left\vert
W_{n-1}\right) -b_{n-1}\left\vert W_{n-2}\right) \right] ,n\geq 2.
\end{eqnarray}%
Here $b_{n}$ are the so-called Lanczos coefficients. They are matrix
elements of Liouville super-operator in the new basis, $b_{n}=\left(
W_{n}\right\vert \mathcal{L}\left\vert W_{n+1}\right) $. We expand the
operator in the Krylov basis as $\left\vert O(t)\right)
=\sum\nolimits_{n}\varphi _{n}(t)\left\vert W_{n}\right) $. The coefficients
$\varphi _{n}(t)$ satisfy the following linear equation,

\begin{equation}
i\partial _{t}\varphi _{n}=-\left( b_{n-1}\varphi _{n-1}+b_{n}\varphi
_{n+1}\right) .
\end{equation}%
Therefore, the operator evolution is mapped to a quantum mechanics problem
of one particle hopping on a tight-binding half-infinite chain. Note that
the $b_{n}$ play the role of the hopping amplitude on this half-infinite
Krylov chain. The particle is initially localized at original site, $n=0$.
During the evolution, it hops into the bulk. The Krylov complexity is
defined as $\left( O(t)\right\vert \mathcal{K}\left\vert O(t)\right)
=\sum\nolimits_{n}n|\varphi _{n}(t)|^{2}$, where $\mathcal{K=}%
\sum\nolimits_{n}n\left\vert W_{n}\right) \left( W_{n}\right\vert $ is the
Krylov super-operator. It is nothing but the center-of-mass of the wave
function on the Krylov chain.

\textit{Models.} To illustrate the Floquet thermalization, we study two
models with periodical driving. One is a quantum spin chain, the other is a
one-dimensional hard-core Bose-Hubbard Model. The evolution operators of
both models in one period are
\begin{equation}
\hat{U}_{\mathrm{F}}^{\alpha }=e^{-i\hat{H}_{\alpha }T/2}e^{-i\hat{V}%
_{\alpha }T/2},  \label{UF}
\end{equation}%
where $\alpha =\mathrm{spin,boson}$, for spin and Bose-Hubbard models
respectively. The corresponding $\hat{H}_{\alpha }$ are
\begin{eqnarray}
\hat{H}_{\mathrm{spin}} &=&\sum_{i=1}^{L}(-J\sigma _{i}^{z}\sigma
_{i+1}^{z}+h_{z}\sigma _{i}^{z}), \\
\hat{H}_{\mathrm{boson}} &=&-\sum_{i=1}^{L}(J\hat{a}_{i}^{\dag }\hat{a}%
_{i+1}+J^{\prime }\hat{a}_{i}^{\dag }\hat{a}_{i+2}+\mathrm{H.c.}).
\end{eqnarray}%
And $\hat{V}_{\alpha }$ are given by $\hat{V}_{\mathrm{spin}}=%
\sum_{i=1}^{L}h_{x}\sigma _{i}^{x}$, $\hat{V}_{\mathrm{boson}%
}=\sum_{i=1}^{L}U\hat{n}_{i}\hat{n}_{i+1}$, where $\hat{n}_{i}=\hat{a}_{i}^{\dag }%
\hat{a}_{i}$. Both models are chaotic.
\begin{figure}[tbp]
\includegraphics[width=\columnwidth]{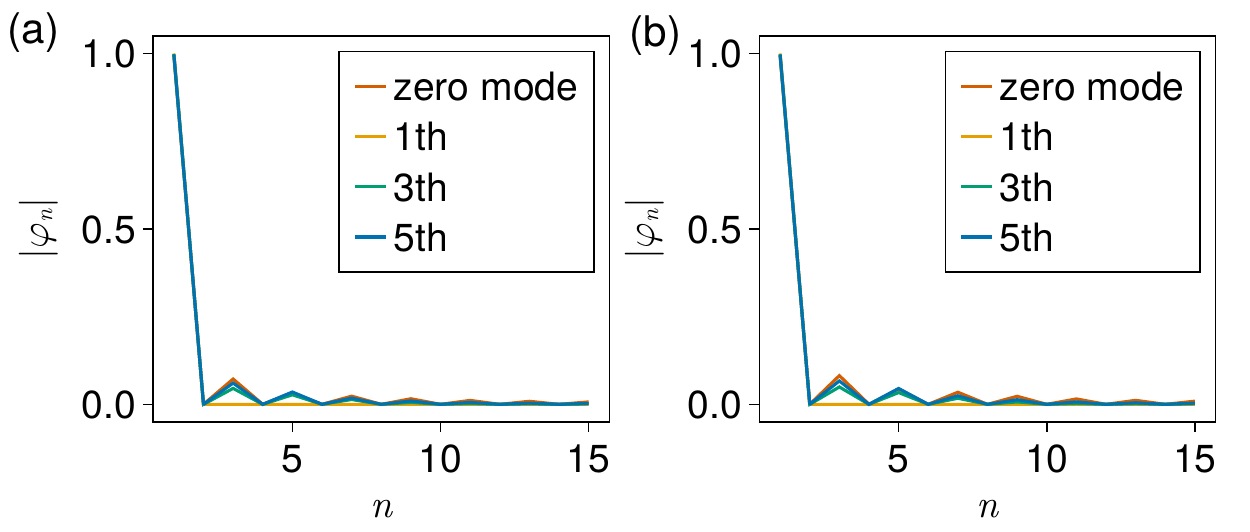}
\caption{Comparison of zero mode and effective Hamiltonian obtained via
Floquet-Magnus expansion in high-frequency regime on Krylov chain. (a)
quantum spin chain at $TJ=1$, (b) hard-core Bose-Hubbard model at $%
TJ=1.3$. For both models, the zero mode is consistent with the
Floquet-Magnus expansion. The simulation parameters are the same as in Fig.%
\protect\ref{fig1}.}
\label{fig2}
\end{figure}

Unlike previous studies~\cite%
{Yates@2021.Krylov-edge,Mitra@2022.F-Krylov,Shrestha@2023.F-Krylov,Suchsland@2023.Krylov, Mitra@2023.F-Krylov}%
, we choose the seed operator as the average Hamiltonian in one period,
\begin{equation}
\hat{O}=\hat{H}^{(0)}=\frac{1}{T}\int\nolimits_{0}^{T}dt\hat{H}(t).
\end{equation}%
For the two models investigated here, we have $\hat{H}_{\alpha }^{(0)}=(\hat{%
H}_{\alpha }+\hat{V}_{\alpha })/2$. We numerically calculate the evolution
operator $\hat{U}_{\mathrm{F}}$ and its logarithm, i.e. $\hat{H}_{\mathrm{eff%
}}^{\mathrm{fold}}$ with finite system size. Then we obtain the Krylov basis
by the method described above. After that, we expand super-operator $%
\mathcal{L}$ on this basis, and obtain the hopping amplitude on the
Krylov chain, $b_{n}$, with varying driving frequencies. Typical
distributions of $b_{n}$ are plotted in Fig.\ref{fig1}(a)(b). We find that,
for both two models, the hopping amplitude fluctuate around a non-zero value
site-by-site in high-frequency driving regime. It reminds us the celebrated
Su-Schrieffer-Heeger (SSH) model, which possesses nontrivial topological band
structure and can sustain ingap edge states. The fluctuation is large near
the origin site, but becomes smaller deep in the bulk. In contrast, in
the low-frequency regime, the fluctuation of hopping amplitude is highly
suppressed, and $b_{n}$ is nearly constant.

We then numerically seek the zero mode operator of Liouvillian, $\mathcal{L}%
\left\vert O\right) =0$. That is equivalent to solving the single-particle eigen
wave function with zero energy on the Krylov chain, see Fig.\ref{fig1}%
(c)(d). We further conclude that the zero mode obtained by this method is
one particular choice of effective Hamiltonian $\hat{H}_{\mathrm{eff}}^{%
\mathrm{zm}}$ with unfolded energy spectrum. The reasons are as follows.
First, this effective Hamiltonian commutes with $\hat{H}_{\mathrm{eff}}^{%
\mathrm{fold}}$, thus is a zero mode of Liouvillian, $\mathcal{L}\hat{H}_{%
\mathrm{eff}}^{\mathrm{zm}}=[ \hat{H}_{\mathrm{eff}}^{\mathrm{fold}},\hat{H}%
_{\mathrm{eff}}^{\mathrm{zm}}] =0$.
Second, we analytically calculate the effective Hamiltonian via
Floquet-Magnus expansion up to the 5th order in the high-frequency regime~%
\cite{Eckardt@2015.Heff}, and compare it with the zero mode on the Krylov
basis. The results are plotted in Fig.\ref{fig2}. Note that for both models,
the zero mode is consistent with the Floquet-Magnus expansion.

\begin{figure}[tbp]
\centering
\includegraphics[width=\columnwidth]{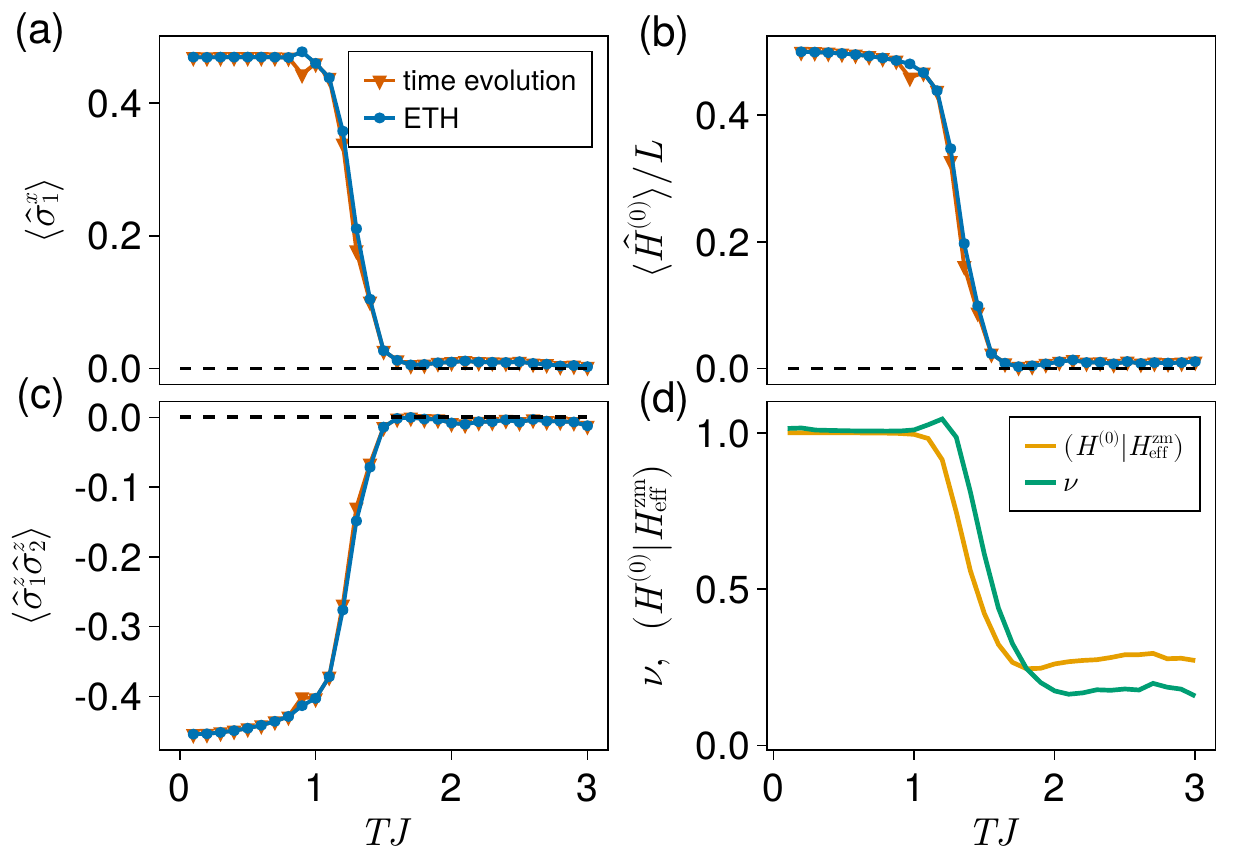}
\caption{Verifying ETH and winding number $\nu$ of Krylov chain in the quantum spin model.
(a-c) Long-time equilibrium values of $\langle\hat{\protect\sigma}%
^x_1(mT)\rangle$, $\langle\hat{\protect\sigma}^z_1\hat{\protect\sigma}%
^z_2(mT)\rangle$ and $\langle\hat{\protect H}^{(0)}(mT)\rangle$, as $m\rightarrow\infty$,
are compared to ETH predictions at different driving
periods. The dashed line marks the infinite-temperature value.
(d) Local winding number $\protect\nu$ of the
Krylov chain is compared with the overlap $(H^{(0)}|H_{\mathrm{eff}}^{%
\mathrm{zm}})$, which exhibits the localization of zero mode on the first Krylov site. We work in zero momentum sector of positive parity.
The initial state is chosen as $Z_2$ state $\mathcal{P}|\uparrow\downarrow\cdots\uparrow\downarrow\rangle$, where $\mathcal{P}$ is the projector onto this sector. The simulation parameters
are the same as in Fig.\protect\ref{fig1}.}
\label{fig3}
\end{figure}

\begin{figure}[tbp]
\centering
\includegraphics[width=\columnwidth]{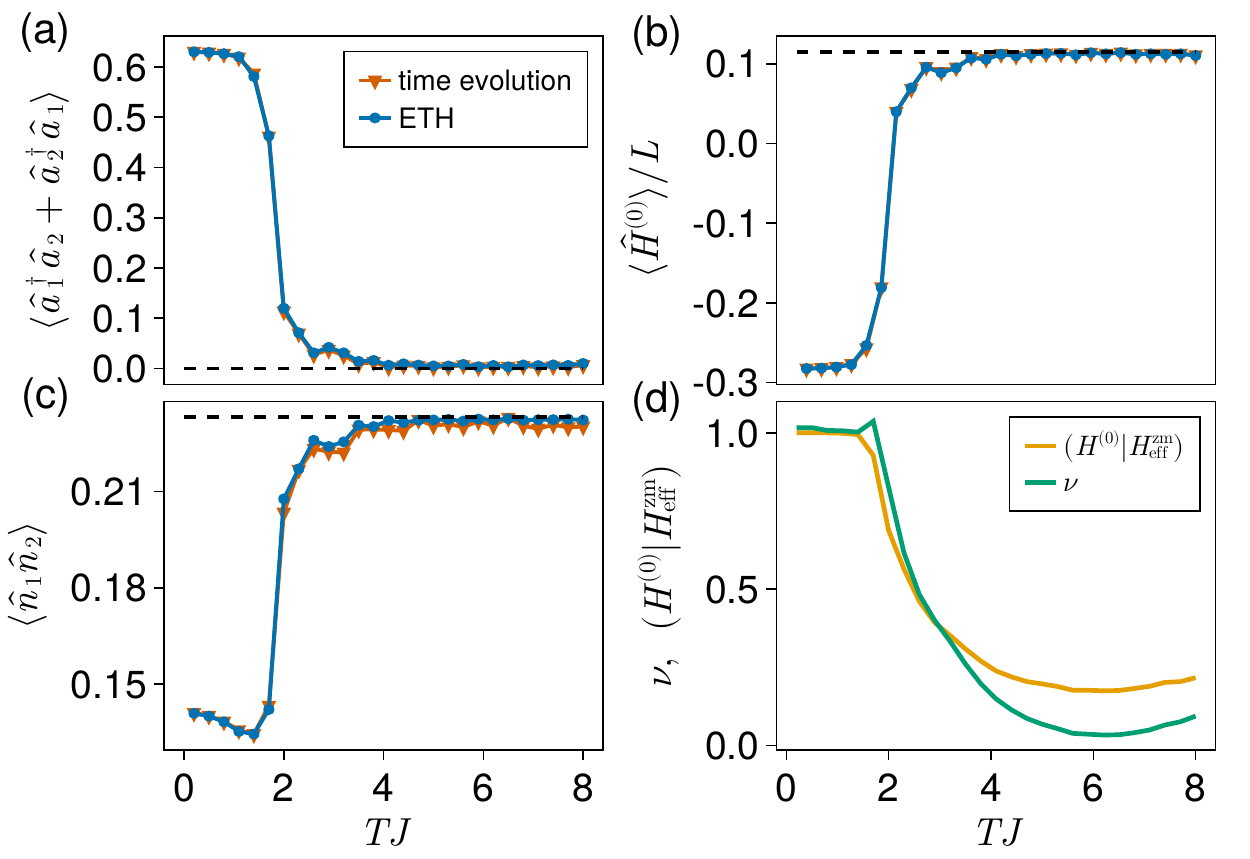}
\caption{Verifying ETH and winding number $\nu$ of Krylov chain in the hard-core Bose-Hubbard
model. (a-c) Long-time equilibrium values of $\langle \hat{b}_1^\dagger
\hat{b}_2(mT) + h.c. \rangle$, $\langle \hat{n}_{1}%
\hat{n}_{2}(mT)\rangle$ and $\langle\hat{\protect H}^{(0)}(mT)\rangle$, as $m\rightarrow\infty$,
are compared to ETH predictions at different driving
periods. The dashed line marks the infinite-temperature value.
(d) Local winding number $\protect\nu$ of the
Krylov chain is compared with the overlap $(H^{(0)}|H_{\mathrm{eff}}^{%
\mathrm{zm}})$, which exhibits the localization of zero mode on the first Krylov site. We work in zero momentum sector of positive parity.
The initial state is chosen as ground state of $\hat{H}_{%
\mathrm{boson}}$. The simulation parameters
are the same as in Fig.\protect\ref{fig1}.}
\label{fig4}
\end{figure}

We note that the distribution of $\hat{H}_{\mathrm{eff}}^{\mathrm{zm}}$ on
the Krylov chain varies as driving period changes, see Fig.\ref{fig1}%
(c)(d). In short driving period regime, the effective Hamiltonian is a edge mode,
localizing\ near the origin of Krylov chain. Thus it possesses small Krylov
complexity. In long driving period regime, $\hat{H}_{\mathrm{eff}}^{\mathrm{zm}}$
extends over the bulk, and has larger Krylov complexity. The behaviors of
the zero mode can be understood by the band topology of the Krylov chain.
Since distribution of $b_{n}$ does not possess the translation symmetry, it is hard to
investigate the topology in momentum space. We turn to calculate the
topological invariance, the winding number, of Krylov chain via a real-space
approach as~\cite%
{Kitaev@2006.Honeycomb,Bianco@2011.RealSpaceTP,Wang@2019.Winding},%
\begin{equation}
\nu =\frac{1}{2N}\mathrm{Tr}\left( \mathcal{SQ}\left[ \mathcal{Q},\mathcal{K}%
\right] \right) .
\end{equation}%
Here $\mathcal{S=}\sum\nolimits_{n}\left( \left\vert W_{2n}\right) \left(
W_{2n}\right\vert -\left\vert W_{2n+1}\right) \left( W_{2n+1}\right\vert
\right) $ is the chiral-symmetry super-operator on the Krylov chain, and $N$
is the lattice length. One note that $\mathcal{SLS}^{\dag }\mathcal{=-L}$,
thus the eigen values\ of $\mathcal{L}$ come in pairs. The super-operator $%
\mathcal{Q}$ is given by $\mathcal{Q=}\sum\nolimits_{i}\left( \left\vert
A_{i}\right) \left( A_{i}\right\vert -\left\vert A_{i}^{\prime }\right)
\left( A_{i}^{\prime }\right\vert \right) $, where $\left\vert A_{i}\right) $
are the eigen modes of $\mathcal{L}$ with positive eigen values $\lambda
_{i} $, $\mathcal{L}\left\vert A_{i}\right) =\lambda _{i}\left\vert
A_{i}\right) $, and $\left\vert A_{i}^{\prime }\right) $ are chiral-symmetry
partners of $\left\vert A_{i}\right) $, $\left\vert A_{i}^{\prime }\right) =%
\mathcal{S}\left\vert A_{i}\right) $. Therefore they have negative eigen
values, $\mathcal{L}\left\vert A_{i}^{\prime }\right) =-\lambda
_{i}\left\vert A_{i}^{\prime }\right) $. We numerically calculate the
winding number $\nu $, and compare it with the overlap $(H^{(0)}|H_{\mathrm{%
eff}}^{\mathrm{zm}})$ as a function of driving frequency. The latter
represents the weight of zero mode on the origin site, and can be used to
characterize the edge state. If $(H^{(0)}|H_{\mathrm{eff}}^{\mathrm{zm}%
})\sim 1$, that indicates the zero mode is well localized on the edge. The
results are plotted in Fig.\ref{fig3}(d) and Fig.\ref{fig4}(d). Note that the
band topology of the Krylov chain determines the zero edge mode. In the short driving period regime, winding number $\nu \approx 1$, that protects a zero edge
mode on the boundary. In long driving period regime, since the gap is closed,
the winding number is not well defined, and gives a non-integer. Thus zero
mode is extended to bulk.

Unlike the folded effective Hamiltonian $\hat{H}_{\mathrm{eff}}^{\mathrm{fold}}$, we found that this zero mode $\hat{H}_{\mathrm{eff}}^{\mathrm{zm}}$ satisfies
the assumption of ETH~\cite{SM}. We can use it to predict the long-time
equilibrium values of local observables as%
\begin{equation}
\left\langle \hat{O}_{\mathrm{local}}(mT\rightarrow \infty )\right\rangle
\approx \mathrm{Tr}\left( \hat{\rho}_{\beta }\hat{O}_{\mathrm{local}}\right)
,
\end{equation}%
where the canonical ensemble density matrix, $\hat{\rho}_{\beta }=e^{-\beta
\hat{H}_{\mathrm{eff}}^{\mathrm{zm}}}/{\mathrm{Tr}}(e^{-\beta \hat{H}_{%
\mathrm{eff}}^{\mathrm{zm}}})$. The inverse temperature $\beta $ is obtained by solving a single equation, $\left\langle \psi (0)\right\vert \hat{H}_{%
\mathrm{eff}}^{\mathrm{zm}}\left\vert \psi (0)\right\rangle =\mathrm{Tr}(%
\hat{H}_{\mathrm{eff}}^{\mathrm{zm}}\hat{\rho}_{\beta })$. Here $\left\vert
\psi (0)\right\rangle $ is a generic initial pure state. In Fig.\ref{fig3}
and Fig.\ref{fig4}, we plot the comparison of local observables after long
time evolution with the ETH prediction. All these calculations start from
the same initial pure state. Note that, for both models, the long time
expectation values are consistent with the ETH prediction not even in high
frequency limit, but in the whole frequency regime. By comparing the long time thermalization behavior with $(H^{(0)}|H_{\mathrm{eff}}^{\mathrm{zm}})$ (Fig.\ref{fig3} and Fig.\ref{fig4}),
we note that when the zero mode is localized on the
edge, system will be driven into an ensemble with finite temperature. If the
zero mode is extended to the bulk, the system will be heated to infinite
temperature. Thus we conclude that it is the band topology of Krylov chain
that determines the long time thermalization of periodically driven many-body
systems. When the Krylov chain is topologically non-trivial, the Floquet
heating will be bounded to finite energy density. When the Krylov chain is
topological trivial or gapless, the heating will be unbounded.

\begin{figure}[tbp]
\centering
\includegraphics[width=\columnwidth]{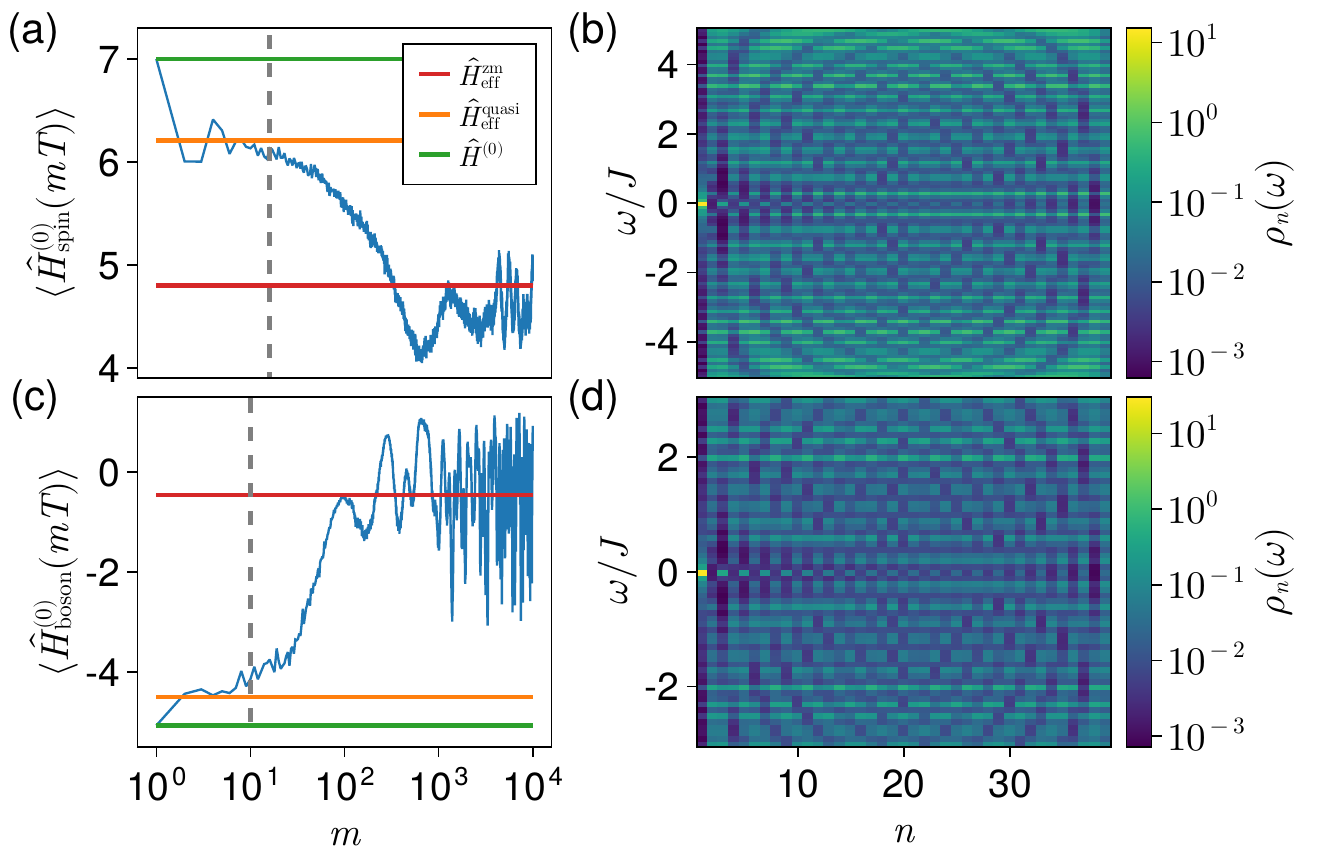}
\caption{Real time thermalization dynamics and local density-of-state on Krylov chains.
(a-b) is the quatnum spin model, with $L=14$ and $TJ=1.2$. (c-d) is the hard-core Bose-Hubbard model  with $L=18$ and $%
TJ=3.9$. (a)(c) The real time thermalization dynamics of local
observable $\langle\hat{H}^{(0)}_{\alpha}(mT)\rangle$ in a long time scale. The predicted values
of ETH given by $\hat{H}_{\mathrm{eff}}^{(0)}$, $\hat{H}_{\mathrm{eff}}^{%
\mathrm{quasi}}$ and $\hat{H}_{\mathrm{eff}}^{\mathrm{zm}}$ are shown by green, yellow and red lines. The dotted gray vertical line marks the value of prethemalization time scale $\protect\tau_*$. (b)(d) Local
density-of-state on Krylov chains. Note that there exists local gap near the origin for both models.}
\label{fig5}
\end{figure}

\textit{Prethermalization.} In the thermodynamic limit, the hopping
amplitude $b_{n}$ at large $n$ approaches a constant. Therefore, the Krylov
chain is gapless, and there is no eigen edge mode. As a result, the system
will be finally heated to infinite temperature despite the frequency.
However, in the high frequency limit, the configuration of hopping amplitude
near the origin site is still similar to the SSH model. That indicates,
there exists local gap near the origin. To see this, we plot the local
density-of-state $\rho _{n}\left( \omega \right) =-\frac{1}{\pi }\mathrm{Im}%
\left\langle W_{n}\right\vert \left( \omega -\mathcal{L}+i0^{+}\right)
^{-1}\left\vert W_{n}\right\rangle $ on the Krylov chain in Fig.\ref{fig5}.
Such local gap will sustain a quasi eigen edge mode, denoted by $\hat{H}_{%
\mathrm{eff}}^{\mathrm{quasi}}$. This quasi edge mode has a finite lifetime,
$\tau _{\ast }$. Before this time scale, the initial wavepacket stays
localized near the edge. But after $\tau _{\ast }$, it will tunnel across
the local gap, and spread to the bulk.

The damping of this quasi edge mode in the Krylov chain describes the
Floquet prethermalization. The lifetime of the quasi edge mode $\tau _{\ast
} $ is just the time scale of prethermalization. During this time scale, the
system will thermalize to a quasi steady state with finite temperature given
by $\hat{H}_{\mathrm{eff}}^{\mathrm{quasi}}$. After that, the system will be
continuously heated to infinite temperature. To see that, we plot the
thermalization process of local observables in a long time scale. We find
that their prethermal values are consistent with the prediction of ETH given
by $\hat{H}_{\mathrm{eff}}^{\mathrm{quasi}}$, but away from the prediction of $\hat{H}_{\mathrm{eff}}^{\mathrm{zm}}$, and $\hat{H}^{(0)}$ see Fig.\ref{fig5}(a)(c).
Numerically, we approximate the $\hat{H}_{\mathrm{eff}}^{\mathrm{quasi}}$ by
truncating the zero mode to the length of local gap regime. Therefore, we
conclude that the Floquet prethermalization process is dominated by the
local topological structure on Krylov chain. With this picture in mind, one
can optimize the prethermal Hamiltonian as the proper quasi edge mode $\hat{H%
}_{\mathrm{eff}}^{\mathrm{quasi}}$.

\textit{Summary.} In summary, we study thermalization of periodically driven
many-body quantum systems via Krylov basis. We found that the zero mode on
the Krylov chain represents the unfolded Floquet effective Hamiltonian. We
demonstrate that the band topology of the Krylov chain dominates the
thermalization behavior. We show that the prethermalization phenomena can be
understood as the damping process of a quasi edge mode protected by the
local gap. Our results bring out the generic connection between Krylov
complexity and thermalization of periodically driven systems that deserves
further studies.

\section*{Acknowledgments}

We thank H. Zhai and Z. Wang for helpful discussion. This work is supported
by NSFC (Grant No. GG2030007011 and No. GG2030040453) and Innovation Program
for Quantum Science and Technology (No.2021ZD0302004).

\end{document}